\documentclass{article}

\usepackage{arxiv}
\usepackage[utf8]{inputenc} 
\usepackage[T1]{fontenc}    
\usepackage{hyperref}       
\usepackage{url}            
\usepackage{booktabs}       
\usepackage{amsfonts}       
\usepackage{nicefrac}       
\usepackage{microtype}      
\usepackage{lipsum}
\usepackage{graphicx}

\title{Analytic self-similar solutions of the Kardar-Parisi-Zhang interface growing equation with various noise terms}

\author{
  Imre Ferenc Barna \\
  Wigner Research Center\\
  Hungarian Academy of Sciences\\
  H-1525 Budapest, P.O. Box 49, Hungary \\
  \texttt{barna.imre@wigner.mta.hu} \\
   \And
 Gabriella Bogn\'ar \\
  Institute of Machine and Product Design\\
  University of Miskolc\\
  Miskolc-Egyetemv\'aros 3515, Hungary \\
  \And
   Mohammed Guedda\\
  Faculte de Mathematiques et d'Informatique\\
  Universit\'e de Picardie Jules Verne Amiens\\
  33, rue Saint-Leu 80039 Amiens, France \\
  \And
  Kriszti\'an Hricz\'o \\
  Institute of Mathematics\\
  University of Miskolc\\
  Miskolc-Egyetemv\'aros 3515, Hungary \\
  \And
  L\'aszl\'o  M\'aty\'as \\
  Department of Technical and Natural Sciences\\
  Sapientia University\\
  Libert\u{a}tii sq. 1, 530104 Miercurea Ciuc, Romania \\
 }

\begin{document}
\maketitle

\begin{abstract}
The one-dimensional Kardar-Parisi-Zhang dynamic interface growth
equation with the self-similar Ansatz is analyzed. As a new feature
additional analytic terms are added. From the mathematical point of
view, these can be considered as various noise distribution
functions. Six different cases were investigated among others
Gaussian, Lorentzian, white or even pink noise. Analytic solutions
were evaluated and analyzed for all cases. All results are
expressible with various special functions like Kummer, Heun,
Whittaker or error functions showing a very rich mathematical
structure with some common general characteristics.
\end{abstract}


\section{Introduction}
Growth patterns in clusters and sodification fronts are challenging
problems from a long time. Basic knowledge of the roughness of
growing crystalline facets has obvious technical applications
\cite{konyv}. The simplest nonlinear generalization of the
ubiquitous diffusion equation is the so called Kardar-Parisi-Zhang
(KPZ) model obtained from Langevin equation
\begin{equation}
\frac{\partial u}{\partial t} = \nu {\bf{\nabla}}^2 u +
\frac{\lambda}{2}({\bf{\nabla}} u)^2 + \eta({\bf{x}},t), \label{kpz}
\end{equation}
where $u$ stands for the profile of the local growth \cite{kpz}. The
first term on the right hand side describes relaxation of the
interface by a surface tension, which prefers a smooth surface.  The
second term is the lowest-order nonlinear term that can appear in
the surface growth equation justified with the Eden model and
originates from the tendency of the surface to locally grow normal
to itself and has a  non-equilibrium in origin. The last term is a
Langevin noise to mimic the stochastic nature of any growth process
and has a Gaussian distribution usually. There are numerous studies
available about the KPZ equation in the literature in the last two
decades. Without completeness we mention some of them. The
foundation of the physics of surface growth can be found in the book
of Barab\'asi and Stanley \cite{barab}. Hwa and Frey
\cite{hwa1,hwa2} investigated the KPZ model with the help of the
self-mode-coupling method and with renormalization group-theory,
which is an exhaustive and sophisticated method using Green's
functions. They considered various dynamical scaling form of $C(x,t)
= x^{-2\varphi} C(bx,b^zt)$ for the correlation function (where
$\varphi, b $ and $z$ are real constants). L\"assig showed how the
KPZ model can be derived and investigated with field theoretical
approach \cite{lass}. In a topical review paper Kriecherbauer and
Krug \cite{krug} derived the KPZ model from hydrodynamical
conservation equations with a general current density relation.
Later, Einax {\it{et al.}} \cite{einax} published a review study on
cluster growth on surfaces.

Numerous models exist, which may lead to similar equations as the
KPZ model, i.e., the interface growth of bacterial colonies
\cite{Matsushita}. More general interface growing models were
developed based on the so-called Kuramoto- Sivashinsky (KS) equation
which is similar to the KPZ model with and extra $ -\nabla^4 u$ term
on the right hand side of (1) (see \cite{kuram1}, \cite{kuram2}).

Guedda has already investigated the generalized deterministic KPZ
equation, when the gradient term is on an arbitrary exponent, with
the self-similar Ansatz \cite{guedda}. Kersner and Vicsek
investigated the traveling wave dynamics of the singular interface
equation \cite{kersner}, which is closely related to he KPZ
equation. \'Odor and co-worker intensively examined the two
dimensional KPZ equation with extended dynamical simulations to
study the physical aging properties of different systems like
glasses or polymers \cite{odor}.

Beyond these continuous models based on partial differential
equations (PDEs) there are numerous purely numerical methods
available to study diverse surface growth phenomena. Without
completeness, we mention the kinetic Monte Carlo \cite{mart},
Lattice-Boltzmann simulations \cite{sergi} and the etching model
\cite{melo}.

By present work, one may find certain kind of solutions to the
problem \cite{CaDoRo10, SaSp10}. It is already mentioned in
\cite{CaDo11} these are for droplet initial conditions.

The first term on the right hand side of the equation can be also
related to diffusion \cite{DoTh17}, and it can be found in the
description of such processes \cite{MaGa05,MaTeVo01}.

In this paper we analyze the solutions of the KPZ equation with
the self-similar Ansatz in one-dimension applying various forms of
the noise term. Numerical results are provided both for similarity
solutions with similarity variables and for the solutions with the
original variables as well. The effect of the parameters involved in the
problem is examined.

The similarity method was used for the investigation of analytic
solution of the two dimensional Navier-Stokes equation with a
non-Newtonian type of viscosity \cite{barna2016self}.

\section{Theory}
Non-linear PDEs has no general mathematical theory, which could help
us to derive physically relevant solutions. Basically, there are two
different trial functions (or Ansatz)  having well-founded physical
interpretation. The first one is the traveling wave solution, which
mimics the wave property of the investigated phenomena described by
the non-linear PDE. The second one is the self-similar Ansatz of the
form
\begin{equation}
u(x,t)=t^{-\alpha}f\left(\frac{x}{t^\beta}\right):=t^{-\alpha}f(\omega),
\label{self}
\end{equation}
where $u(x,t)$ can be an arbitrary variable of a PDE and $t$ denotes
time and $x$ means spatial dependence. The similarity exponents
$\alpha$ and $\beta$ are of primary physical importance since
$\alpha$ represents the rate of decay (or sharpening process if
$\alpha <0 $) of the magnitude $u(x,t)$, while $\beta$  is the rate
of spread (or contraction if  $\beta<0$) of the space distribution
for $t > 0 $. The most powerful result of this Ansatz is the
fundamental or Gaussian solution of the Fourier heat conduction
equation (or for Fick's diffusion equation) with $\alpha =\beta =
1/2$. These solutions are exhibited on Fig. 1 for fixed times
$t_1<t_2$. We can generally state, that this Ansatz mimics the
diffusive properties (the similarities to normal diffusion) of the
investigated  PDE. This is the key point why this Ansatz is used. We
note, that in some cases \cite{barn} the traveling wave and
self-similar solutions are intertwinted and can be transformed into
one another.

This transformation is based on the assumption that a self-similar solution exists, i.e., every
physical parameter preserves its shape during the expansion.
Self-similar solutions usually describe the asymptotic behavior of
an unbounded or a far-field problem; the time $t$ and the spacial
coordinate $x$ appear only in the combination of  $f(x/t^{\beta})$.
It means that the existence of self-similar variables implies the
lack of characteristic lengths and times. These solutions are
usually not unique and do not take into account the initial stage of
the physical expansion process. It is also transparent from
(\ref{self}) that to avoid singularity at $t=0$  the following
transformation $\tilde{t} = t+t_0$ is valid.

We should note that with the application of the Hopf-Cole transformation  $ h = A \ ln(y)$
the non-linear KPZ equation can be converted to the regular heat conduction (or diffusion equation) with a stochastic term.

There are numerous reasonable generalization of (\ref{self})
available, one of them is $u(x,t) = h(t)\cdot f[x/g(t)] $, where
$h(t)$ and $g(t)$ are continuous functions. The choice of $h(t)=g(t)
= \sqrt{t_0-t}$  is a special kind, called the blow-up solution. It
means that the solution becomes infinity after a well-defined finite
time duration.

These kind of solutions  describe the intermediate asymptotics of a
problem. They hold when the precise initial conditions are no longer
important, but before the system has reached its final steady state.
For some systems it can be shown that the self-similar solution
fulfills the source type (Dirac delta) initial condition. They are
much simpler than the full solutions and so easier to understand and
study in different regions of parameter space. A final reason for
studying them is that they are solutions of a system of (ordinary
differential equations (ODEs) and they do not suffer the extra
inherent numerical problems related to  PDEs. In some cases
self-similar solutions helps us to understand global physical
properties of the solutions like finite oscillations, diffusion-like
properties, discontinuous solutions or the existence of compact
supports. Such kind of general information is hard to find from
purely numerical calculations.

Applicability of this Ansatz is quite wide and comes up in various
mechanical systems \cite{sedov}, in transport phenomena like heat
conduction \cite{ barn}, in Euler equation \cite{barna2013analytic}
or even in various two or three dimensional Navier-Stokes equations
\cite{barna2,barna3}.

\section{Results without noise term}\label{s:3}

\begin{figure}
	\begin{center}
		\includegraphics[keepaspectratio, width=7cm]{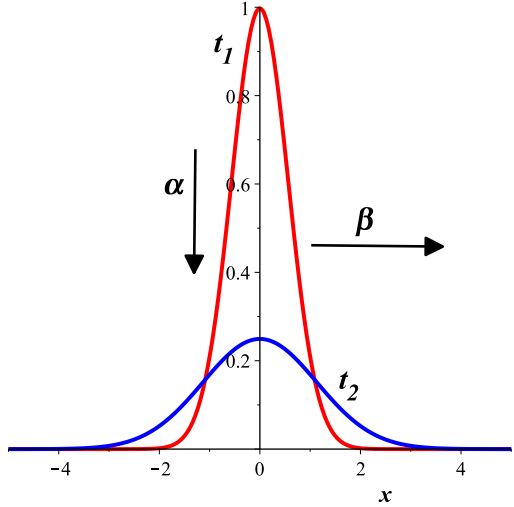}
		\caption{A self-similar solution of Eq. (\ref{self}) for $t_1<t_2$.
			The presented curves are Gaussians for regular heat conduction.}
		\label{egyes}       
	\end{center}
\end{figure}
We start our investigation with the KPZ equation (\ref{kpz}) in one
spatial dimension neglecting the noise term ($\eta(x,t)=0$).
Calculating the  time and spacial derivatives of  (\ref{self}) and
substituting to (\ref{kpz}) one gets the following constrains for
the exponents: $\alpha = 0$ and $\beta = 1/2$. In regular heat
conduction (or diffusion) process both exponents are equal to $1/2$,
which means that the decay (perpendicular dynamics to the surface)
and the spreading (parallel dynamics to the surface) of the solution
have the same strength in time. For the KPZ equation the general
features are different. Now, $\alpha$ vanishes, which means that we
cannot identify any kind of decaying dynamics of the solution
perpendicular to the surface. The non-zero value of $\beta$ can be
understood as a kind of spreading parallel to the surface. These are
general and relevant statements of the surface growth process
described by our solution. The remaining non-linear ODE reads
\begin{equation}
\nu f''(\omega) + f'(\omega)\left [\frac{\omega}{2}+ \frac{\lambda}{2} f'(\omega)\right] = 0.
\label{ode}
\end{equation}
The general solution can be given with the logarithm of the error
function
\begin{equation}
f(\omega) = 2 \frac{\nu  } {\lambda} ln \left( \frac{\lambda c_1
	\sqrt{\pi\nu} \> \> erf[\omega/(2\sqrt{\nu})] + c_2}{2\nu}  \right)
, \label{sol}
\end{equation}
where $ erf$ is the error function \cite{NIST} and $c_1$ and $c_2$
are integration constants. The role of $c_2$ is just a shift of the
solution. For physical reasons  the surface tension $\nu$ should be
larger than zero. Analyzing the solution to Eq. (\ref{sol}), the
value of $ \lambda$ should be positive as well. Figure 2 presents
three different shape function solutions of the ODE with $c_1 = c_2
= 1$ and for three different  combinations of $ \lambda $ and $\nu$.
Note, that all solution has the same simple qualitative behavior, a
quick ramp-up and a converged plateau. Figure 3 shows the complete
solutions of the original PDE showing the spatial and time
dependence for $c_1=c_2=\lambda=\nu =1$. The function has a similar
structure like Fig. 2 a quick ramp-up and a slow convergent plateau.
We may say that different numerical values of $\eta$ and $\nu$ do
not drastically change the qualitative structure of the solution
surface. A closer look of the solution shows, that at $ t = 0 $ the
height of the surface has a constant value, later at small times a
thin valley is formed, which becomes wider and wider as time goes
on. The physical parameters $\lambda, \nu$ and the integration
constants (only shifts the solutions) set the shape and the depth of
the valley. Even at large times at large spatial distances, the
height of the surface, i.e., the asymptotic solution remains
constant. The growth of the valley (or void) can be understood as a
kind of front propagation  as well, and therefore can be explained
with the non-zero $\beta$ exponent.

At this point is comes clear to us, that any kind of surface growth mechanism can be only described
and investigated with the additional noise term. The direct application of the self-similar solution to
the KPZ equation without any additional noise term  $\eta(x,t)$ cannot describes any kind of growth process.

\begin{figure}
	\begin{center}
		\includegraphics[keepaspectratio, width=7cm]{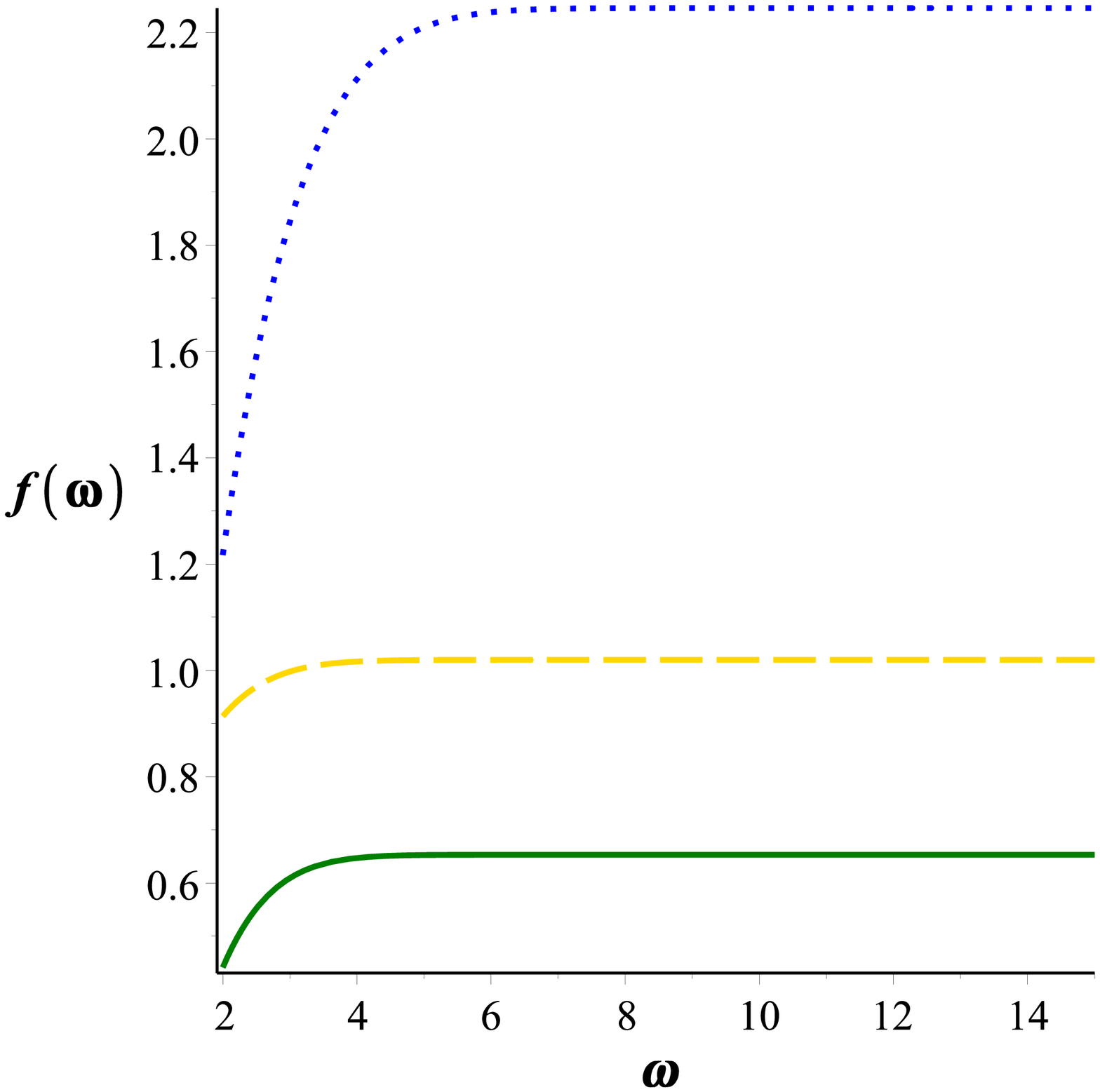}
		\caption{The self-similar solution of the KPZ equation without any
			noise term  for $c_1=c_2 =1$. Solid line is for $\lambda = \nu =1$,
			dashed line is for
			$\nu =1$ and $\lambda =2$ and the dotted line represents $\nu =2$ and $\lambda =1$.}
		\label{kettes}       
		
		\includegraphics[keepaspectratio, width=7cm]{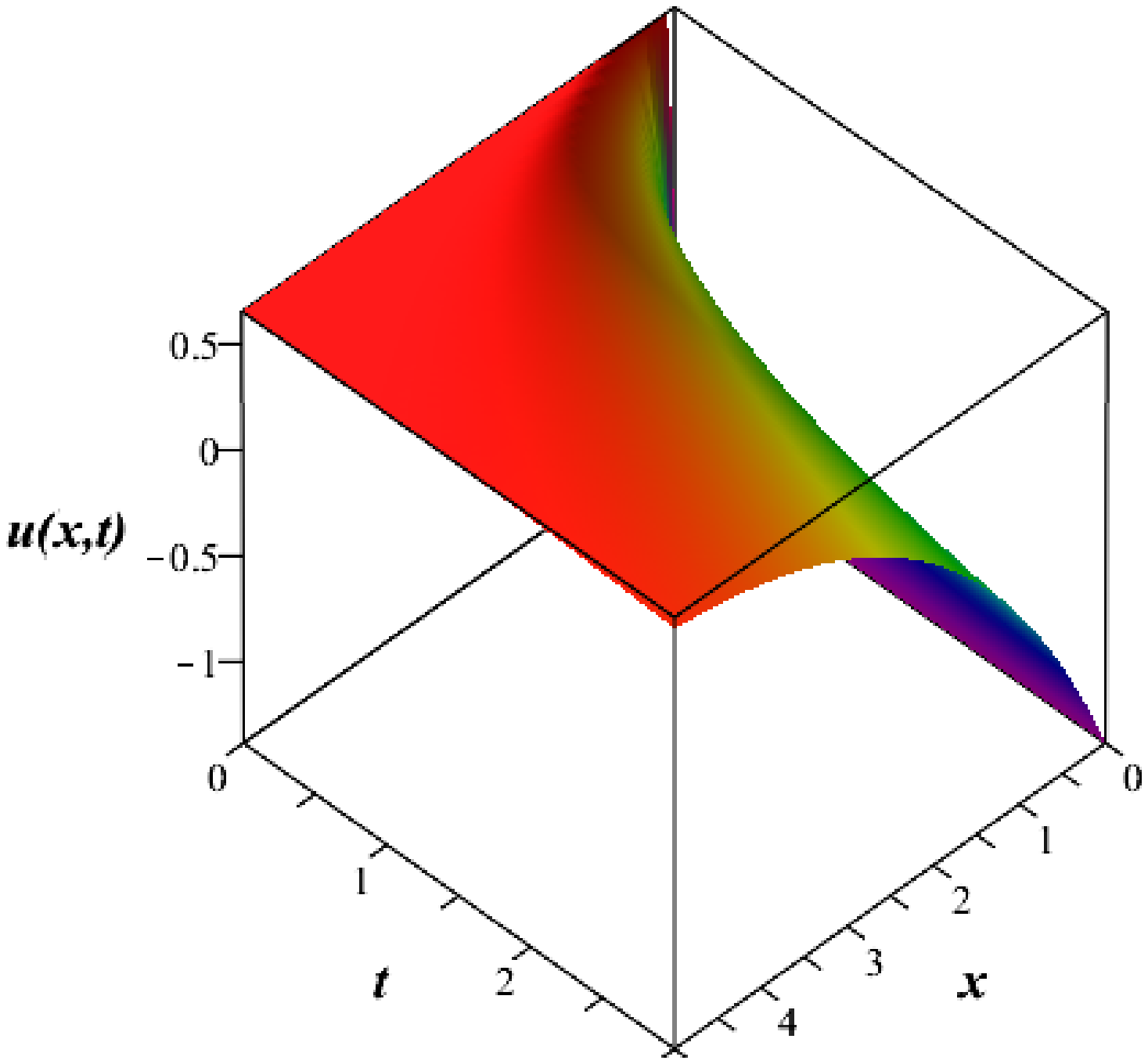}
		\caption{The self-similar solution of the original KPZ equation
			without the noise term for the parameter set $c_1 = c_2 = \lambda =
			\nu =1$ .}
		\label{harmas}       
	\end{center}
\end{figure}

\section{Results with various noise terms}\label{s:4}
It is obvious that every physical process is perturbed with some
kind of perturbations. Perturbations which carry no information are called noise.
The KPZ equation, very correctly, include an additive noise term. Our similarity
Ansatz of the from (\ref{self}) satisfy the general ODE with the form of
\begin{equation}
\nu f''(\omega) + f'(\omega)\left [\frac{\omega}{2}+
\frac{\lambda}{2} f'(\omega)\right]+t \eta (\omega)= 0.
\label{ode_noise}
\end{equation}
If we want to apply the self-similar Ansatz (2) to the noisy KPZ
equation than the noise term  $t \eta(x,t) = l(\omega) =
l(x/t^{\beta}) $  should be some kind of analytic function of the
original variables of $x,t$,  on the other side, we want to handle
noise in a statistically correct manner, $\eta(x,t)$ should be a
density function of a probability distribution as well. This second
condition dictates that the density function should be positive and
should have an existing finite integral on a finite or infinite
support. Note, the extra time dependence of the last term in
(\ref{ode_noise}) is dictated from a dimensional analysis reason.

First, we investigate noises with various power-law dependencies $
l(\omega) = a\omega^n$. Noises with different integer power values
of $n$ are named after different colors $n = -2, -1, 0, 1$ which are
brown, pink, white and blue, respectively. Two additional cases, the
Gaussian and Lorenzian noises, are investigated as well. To avoid
further misunderstanding we must state that in our calculations a
Gaussian noise means that the noise term explicitly depends on the
scaled spatial coordinate $x/t^{1/2}$ and not on the Fourier spectra
as usually considered. The argument $\omega$ of the shape function
is the time-scaled spatial coordinate and not the angular frequency.
Of course, in principle it is possible to evaluate the Fourier
spectra of our noise terms and interpret them in the frequency
domain but that is not the aim of the present study.


\subsection{Brown noise $ n = -2$ }
Our first case leads to the ODE of
\begin{equation}
\nu f''(\omega) + f'(\omega)\left [\frac{\omega}{2}+
\frac{\lambda}{2} f'(\omega)\right]+\frac{a}{\omega^2}= 0.
\label{kpz_egy_per_etanegyzet}
\end{equation}

Using the  mathematical program package Maple 12, the solution can
be obtained in a closed form
\begin{eqnarray}
f(\omega) = -\frac{\omega^2}{4\lambda} + \frac{1}{\lambda}  \left. \bigg[  ln (
( \omega^3  \lambda^2  \{  c_1 M_{-\frac{1}{4},\frac{d}{4}}(r) - c_2 W_{-\frac{1}{4},\frac{d}{4}}(r) \}^2  )/  \right. \nonumber \\
( M_{\frac{3}{4},\frac{d}{4}}(r) \cdot  \nu  \cdot W_{-\frac{1}{4},\frac{d}{4}}(r)  +
M_{\frac{3}{4},\frac{d}{4}}(r) \cdot \nu \cdot d \cdot   W_{-\frac{1}{4},\frac{d}{4}}(r) + \left. \right. \nonumber \\
4M_{-\frac{1}{4},\frac{d}{4}}(r) \cdot \nu \cdot W_{\frac{3}{4},\frac{d}{4}}(r) )^2 ) \nu
\left. 1  \right. \bigg],
\label{zaj1}
\end{eqnarray}
where $M$ and $W$ are the Whittaker $M$ and Whittaker $W$ functions
\cite{NIST}. For the better transparency we used the following
notations  $d=\sqrt{\nu^2-2\lambda a}/{\nu}$ for the second
parameter and $r = \omega^2/(4\nu) $ for the argument of the
Whittaker functions. Both parameters of the Whittaker functions must
be real numbers, which means   that $ \nu^2-2\lambda a \ge 0 $
therefore for any kind of fixed and positive $\nu$ and $\lambda$,
there is an upper limit for $a$, which is the strength of the noise
term. So, if the magnitude of the noise reaches a definite level,
the Whittaker function and the solution of the problem becomes
undefined and meaningless. This is consistent with our physical
picture about noisy processes.

Due to the Whittaker function, the solution is undefined for
negative  arguments $\eta$ for any kind of parameter set. Figure 4
presents the solution of Eq. (\ref{zaj1}) for two different
parameter sets. The numerical value of $\nu$ defines the position of
the singularity in a non trivial way, larger $\nu$ shifts the
position to larger arguments. At fixed physical parameters  $a$,  $
\nu$ and $\lambda$, the first integration constant  $c_1$ is equal
with the asymptotic value of the solution for large arguments. The
second integration constant $c_2$ directly defines the function in
the origin in a non-trivial way, the larger the value the larger the
function as well.

Figure 5 shows the solution profile $u(x,t)$ of the KPZ equation as
the function of time and spatial coordinate. The sharp cusp is
clearly seeing. It is also evident that the position of the cusp
moves to larger spatial coordinates as time goes on, compared to the
free KPZ solution of Eq. (3.1), which means that the additional
noise term puts a small island into the origin which is growing and
pushing the cusp before. Another interesting feature is that the
cusps survives even at large times and is not filled up as would we
expect from our physical intuition.

The sharp but finite cusp that arises in this solutions is the
so-called Van Hove singularity, which was first seen in crystals in
the function of elastic frequency distribution \cite{vanhove}. We
note, that the finite number of peaks on Fig. 5 under the cusp are
just an artifact of the finite resolution of the Maple software.
\begin{figure}
	\begin{center}
		\includegraphics[keepaspectratio, width=7cm]{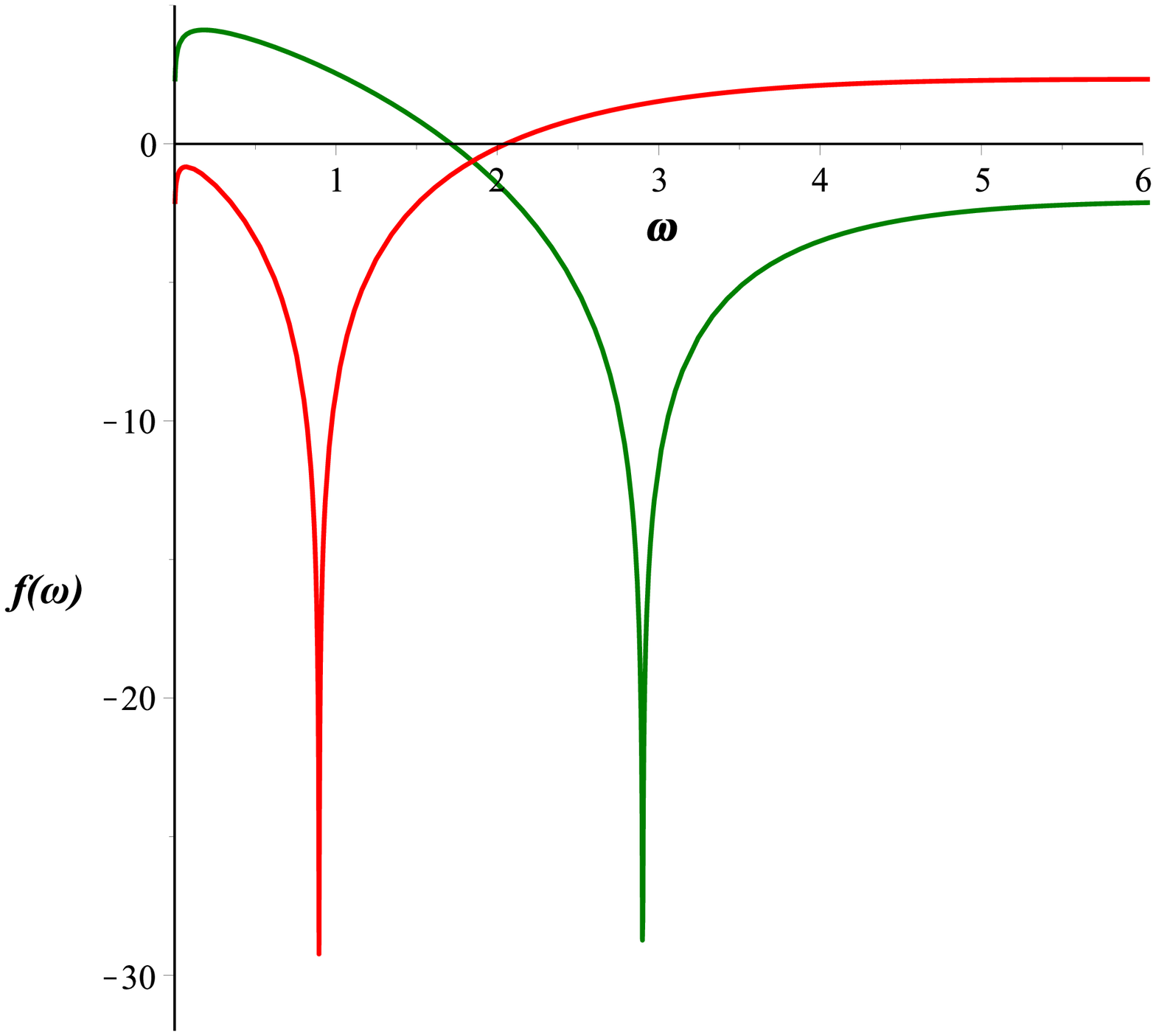}
		\caption{The shape functions of the Brownian noise  Eq. (\ref{zaj1})
			for $a=\lambda = 1$ , $\nu =2$ physical parameters . The red line is
			for  $c_1=3, c_2 =1$ and and the green line is for  the  integration
			constants $c_1=1, c_2 =3$.   }
		\label{nana}       
		
		\includegraphics[keepaspectratio, width=7cm]{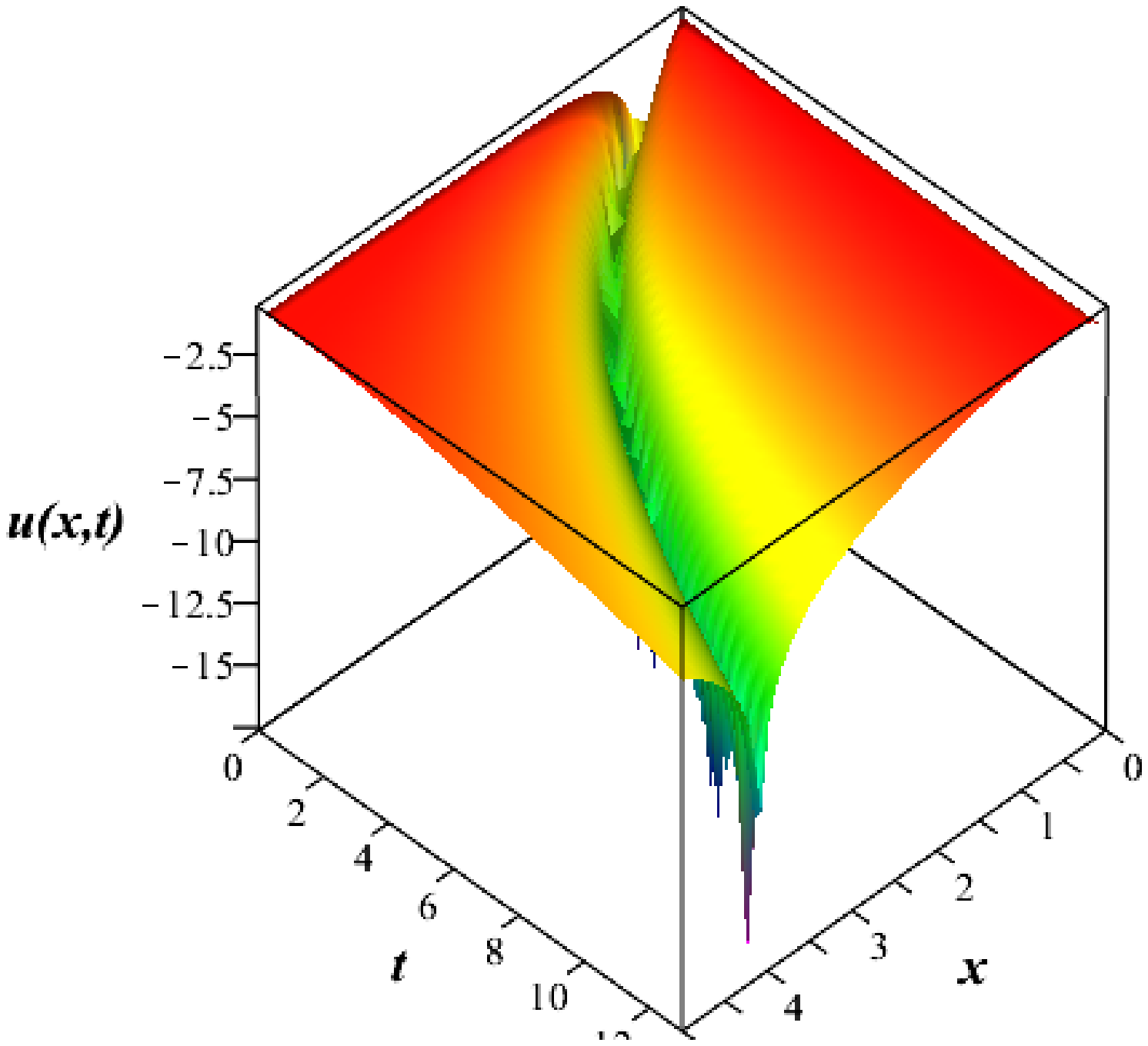}
		\caption{The solution of the KPZ eq. with Brownian noise,  for the
			parameter set of  $ \lambda = \nu = c_1 = c_2 = 1$ and $a=1/2$. }
		\label{harmgdfg}       
	\end{center}
\end{figure}

\subsection{Pink noise $n = -1$}

The corresponding ODE reads
\begin{equation}
\nu f''(\omega) + f'(\omega)\left [\frac{\omega}{2}+
\frac{\lambda}{2} f'(\omega)\right]+ \frac{a}{\omega}= 0 .\label{kpz_const}
\end{equation}

In the most general case when all three parameters are undefined
($\lambda, \nu, a$), there is no closed formula available for the
solution. There is an existing expression containing the integral of
the HeunB functions \cite{NIST} together with other functions. For
given values ($\lambda = \nu = a =1$),  the formula becomes a bit
more transparent
\begin{eqnarray}
f(\eta) = -\frac{1}{\eta^2} + 2ln \left[ \frac{c_1\eta}{2} H_B \left(1, 0, -1, 2,
\frac{-\eta}{2} \right) - \frac{c_2 \eta}{2}H_B \left(1, 0, -1, 2, \frac{-\eta}{2} \right) \times \right.  \nonumber \\
\left.  \left\{ \int \frac{e^{-\frac{\eta^2}{4}}   }{ \eta^2 H_B
	\left(1,0,-1,2, \frac{-\eta}{2} \right)^2}  \right\}  \right]. \ \
\label{egy_per_eta}
\end{eqnarray}

Unfortunately, if the strength of the noise $a$ is different, then
the term  containing the integral of the  function $H_C$ cannot be
separated from the pure $H_C$ function and the final from cannot be
evaluated numerically.

Figure 6 shows the shape function where the physical parameters are
set to unity. At first sight, the solution looks the same as the
solution without noise, however there is a small positive island,
which is created next to the valley. In other words the solution has
a local maximum at finite $\omega$, which means a finite
time and space coordinate $x/t^{\frac{1}{2}}$. As time goes on the
valley becomes wider and wider pushing this tiny island to the right
with a continuous smearing. We can say that the surface growing
phenomena breaks down in the dynamics of this tiny island or rather
"reef". So, there is no general surface growth phenomena along the
whole axis.

Figure 7 shows the complete solution of the KPZ equation with the pink
noise, note the tiny positive bump at small time and space coordinates.

\begin{figure}
	\begin{center}
		\includegraphics[keepaspectratio, width=7cm]{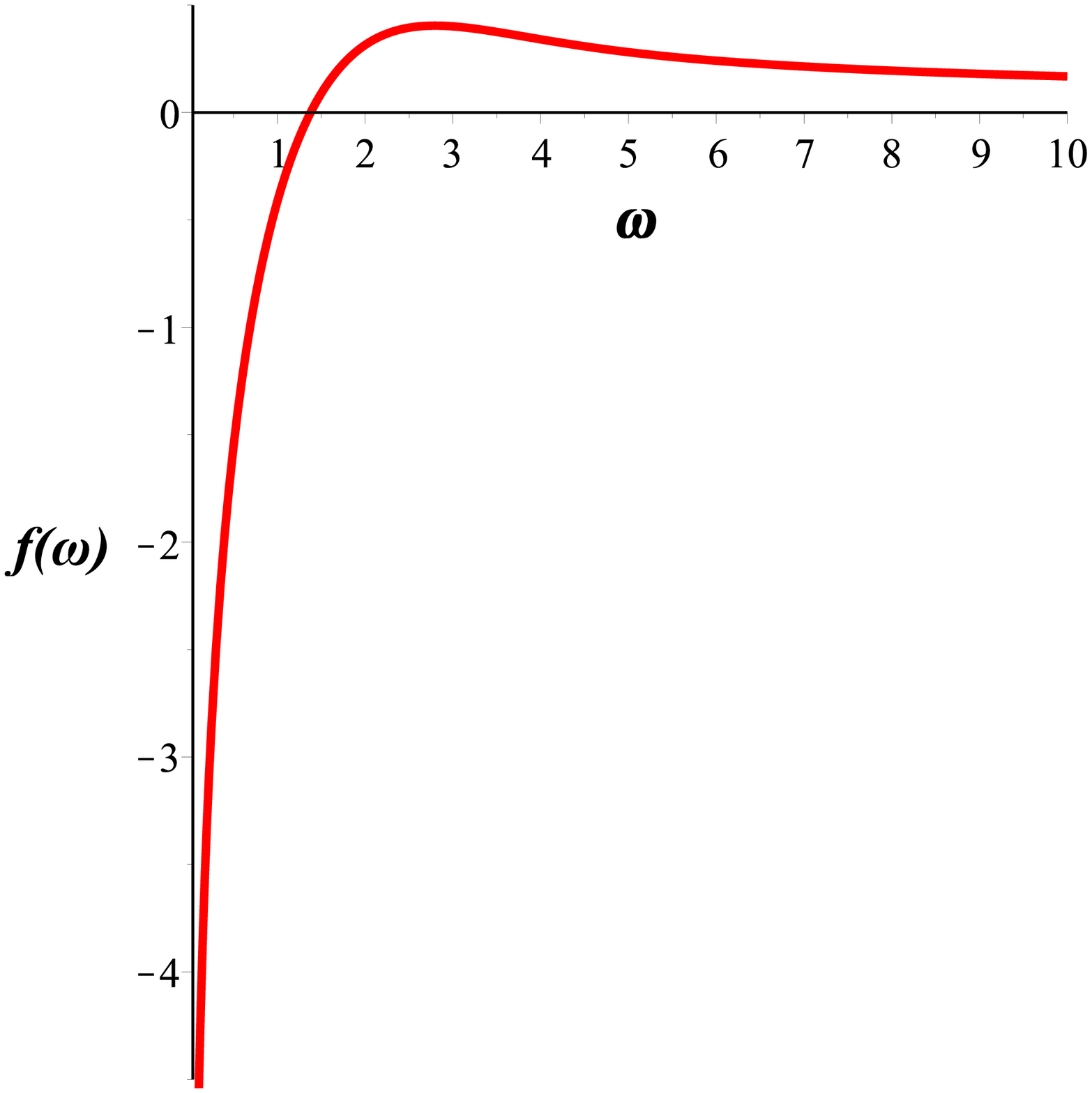}
		\caption{The shape function of the pink noise for $a=
			\lambda = \nu = 1$ and for $c_1 =1 $ and $ c_2 = 0.$ }
		\label{egy_per_omega}       
		
		\includegraphics[keepaspectratio, width=7cm]{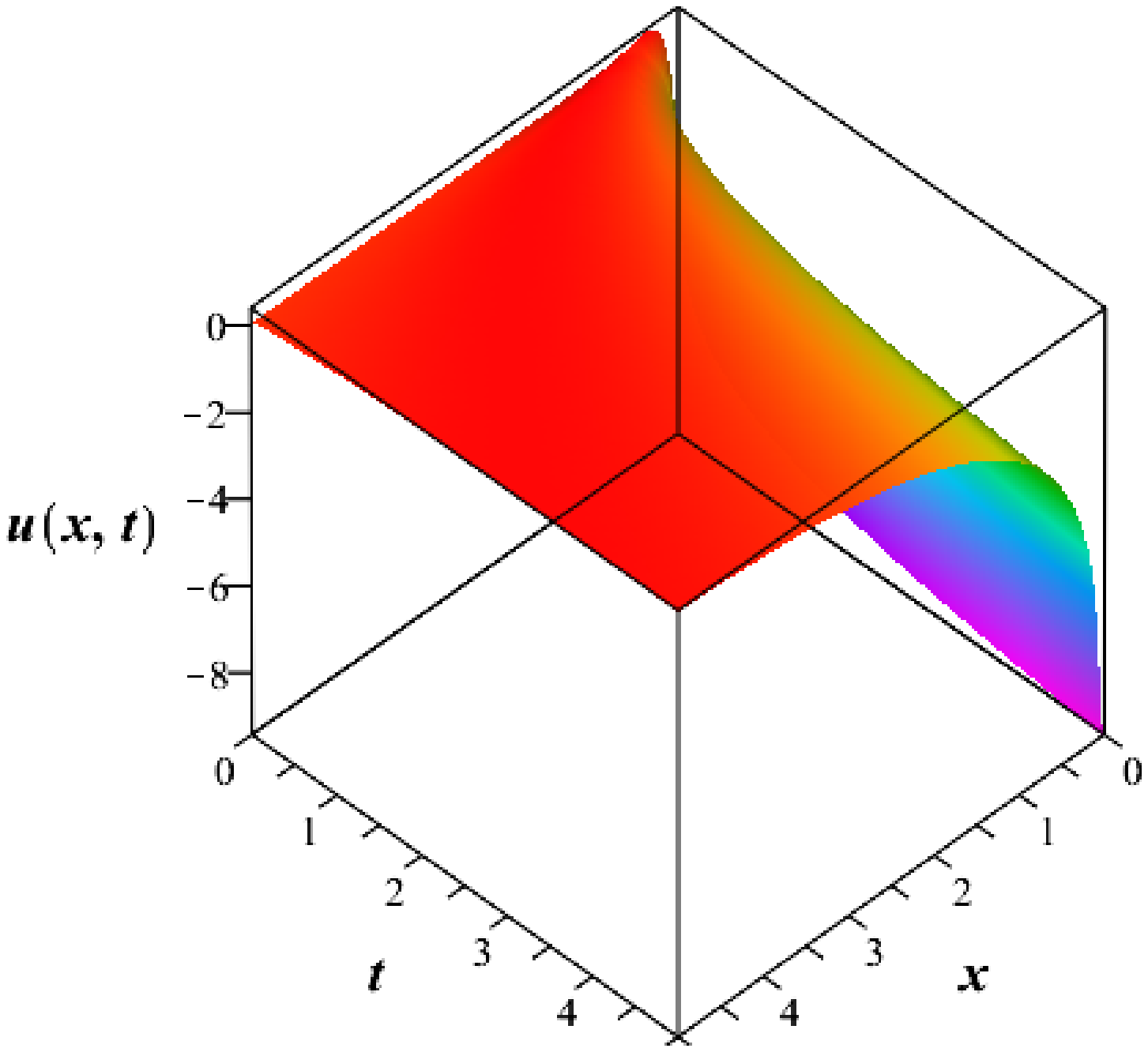}
		\caption{The complete solution of the KPZ equation for the pink  noise with the
			parameters given above. }
		\label{egy_per_omega}       
	\end{center}
\end{figure}

\subsection{White noise $ n = 0 $ }
The associated ODE is now
\begin{equation}
\nu f''(\omega) + f'(\omega)\left [\frac{\omega}{2}+
\frac{\lambda}{2} f'(\omega)\right] + c = 0. \label{kpz_const}
\end{equation}
There is no general closed formula available for a general real
constant $c$. However, if the constant noise term is written in the
form of $\eta = n \lambda$, which can be identified as a  kind of
external mechanism  due to the work of \cite{vicsek}, then other
analytical solutions become available which can be expressed via
Kummer $M$ and Kummer $U$ functions \cite{NIST}

\begin{eqnarray}
f(\omega) =       \frac{2\nu}{\lambda} \ln \left( \left\{ {\lambda}  \left[
-c_1M \left(1-\frac{n\lambda}{2},\frac{3}{2},\frac{\omega^2}{n\lambda}\right) -
c_2 U\left(1-\frac{n\lambda}{2},\frac{3}{2},\frac{\omega^2}{n\lambda} \right) \right]    \right\}   \right.  /   \nonumber \\
\left.  \left\{\nu \left[  M \left(-\frac{n\lambda}{2},\frac{3}{2},\frac{\omega^2}{n\lambda}\right)  U \left(1-\frac{n\lambda}{2},\frac{3}{2},
\frac{\omega^2}{n\lambda}\right) + \right. \right. \right.  \nonumber \\
\left. \left. \left.
n\lambda   M \left(-\frac{n\lambda}{2},\frac{3}{2},\frac{\omega^2}{n\lambda}\right)  U \left(1-\frac{n\lambda}{2},\frac{3}{2},
\frac{\omega^2}{n\lambda}\right)   + \right. \right. \right.  \nonumber \\
\left. \left. \left.
2M \left(1-\frac{n\lambda}{2},\frac{3}{2},\frac{\omega^2}{n\lambda}\right)
U \left(-\frac{n\lambda}{2},\frac{3}{2},
\frac{\omega^2}{n\lambda}\right)   \right]  \right\}  \right) -   \frac{2\nu \ln (2)}{\lambda}
\label{const}.
\end{eqnarray}

\begin{figure}
	\begin{center}
		\includegraphics[keepaspectratio, width=7cm]{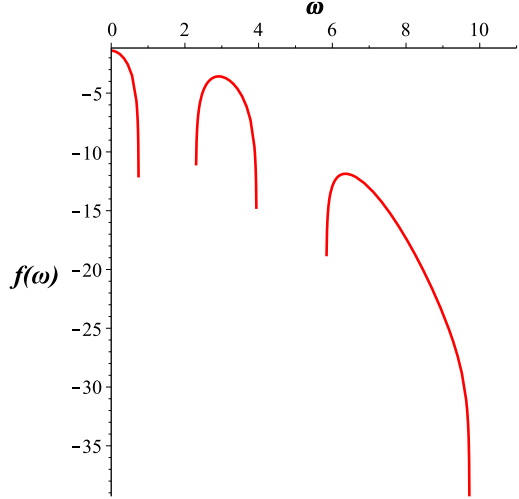}
		\caption{The shape function for the white noise for  parameter set
			$a=\lambda =\nu = 1 =c_1 = c_2 = 1$.}
		\label{const}       
	\end{center}
\end{figure}

\begin{figure}
	\begin{center}
		\includegraphics[keepaspectratio, width=7cm]{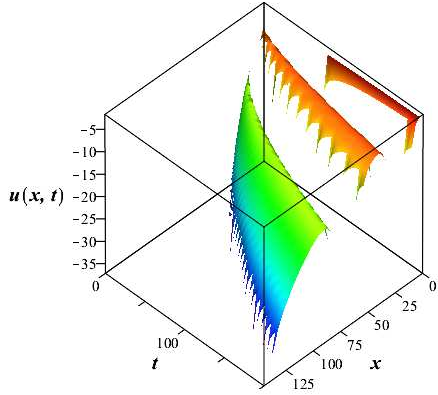}
		\caption{The complete solution for the white noise for $a= 1/2, \lambda = 2$, $\nu = 1/2$ and for
			$c_1 = c_2 = 1.$  }
		\label{const_3d}       
	\end{center}
\end{figure}
Figure  8 shows the shape function for the white noise. The new
feature is that the solution fell apart to numerous distinct
intervals with compact supports. The function has large but finite
negative values at the supports with infinitely large derivatives,
which can be called cusps as well.  Note, that there are finite
intervals where the solution is not defined. Figure 9 shows the
solution $u(x,t)$ of the KPZ equation. We mention that the separate
islands continuously grow as time goes on, but, they cannot touch
each other even at large times. The finite number of peaks under the
cusp are again an artifact of the finite resolution of the Maple
software. Such kind of surface growth, where separate "barrys" are
created, can be noticed on coral reefs or on dripstones.

\subsection{Blue noise $ n = 1 $ }

The last  power law noise case is the following
\begin{equation}
\nu f''(\omega) + f'(\omega)\left [\frac{\omega}{2}+
\frac{\lambda}{2} f'(\omega)\right]+ a\omega =0. \label{kpz_omega}
\end{equation}

The structure of the solution shows some similarity to the brown and
white noise and can be expressed with the help of the Kummer $U$ and
$M$ functions

\begin{eqnarray}
f(\omega) =-2\,\omega+2\,\frac {\nu}{\lambda}\ln  \left\{
\frac{ -\lambda (4\lambda -\omega) \left[ c_2 U(\epsilon_-, \frac{1}{2},\sigma) - c_1 M(\epsilon_-, \frac{1}{2},\sigma)  \right] }
{ 4 \left[ M(\epsilon_+, \frac{1}{2}, \sigma) U(\epsilon_-, \frac{1}{2}, \sigma) \lambda^2 +
	M(\epsilon_1, \frac{1}{2}, \sigma) U(\epsilon_+, \frac{1}{2}, \sigma) \nu
	\right]} \right\}.
\end{eqnarray}
For the better transparency, we use the following notations  $
\epsilon_- = \frac {2\,{\lambda}^{2}-\nu}{\nu}$ ,   $ \epsilon_+ =
\frac {2\,{\lambda}^{2}+\nu}{\nu}$ and $\sigma = {\frac { \left(
		4\,\lambda-\omega \right) ^{2}}{4\nu}} $. Figure 10 presents the
shape function for the blue noise. It shows some similar features to
the former white noise. The solution on the positive axis can be
interpreted only on two separate finite intervals. The function has
finite values, however, the first derivatives at the right hand side
of the intervals become infinite, which can be interpreted as a kind
of "semi-cusp". With some vertical shifts parallel to the
axis $f(\omega)$ the solution can be physically interpreted as two
distinct islands  growing in  time. As an additional fineness, we
note that at left side of the right island there is a gap.

Figure 11 shows the solution function $u(x,t)$ of the original PDE.
The spatial range of the two distinct intervals is continuously
growing in time, however, it remains separate even at infinite
times. With this kind of noise and Ansatz, there is no way to grow a
constant surface above the whole positive semi-axis.

\begin{figure}
	\begin{center}
		\includegraphics[keepaspectratio, width=7cm]{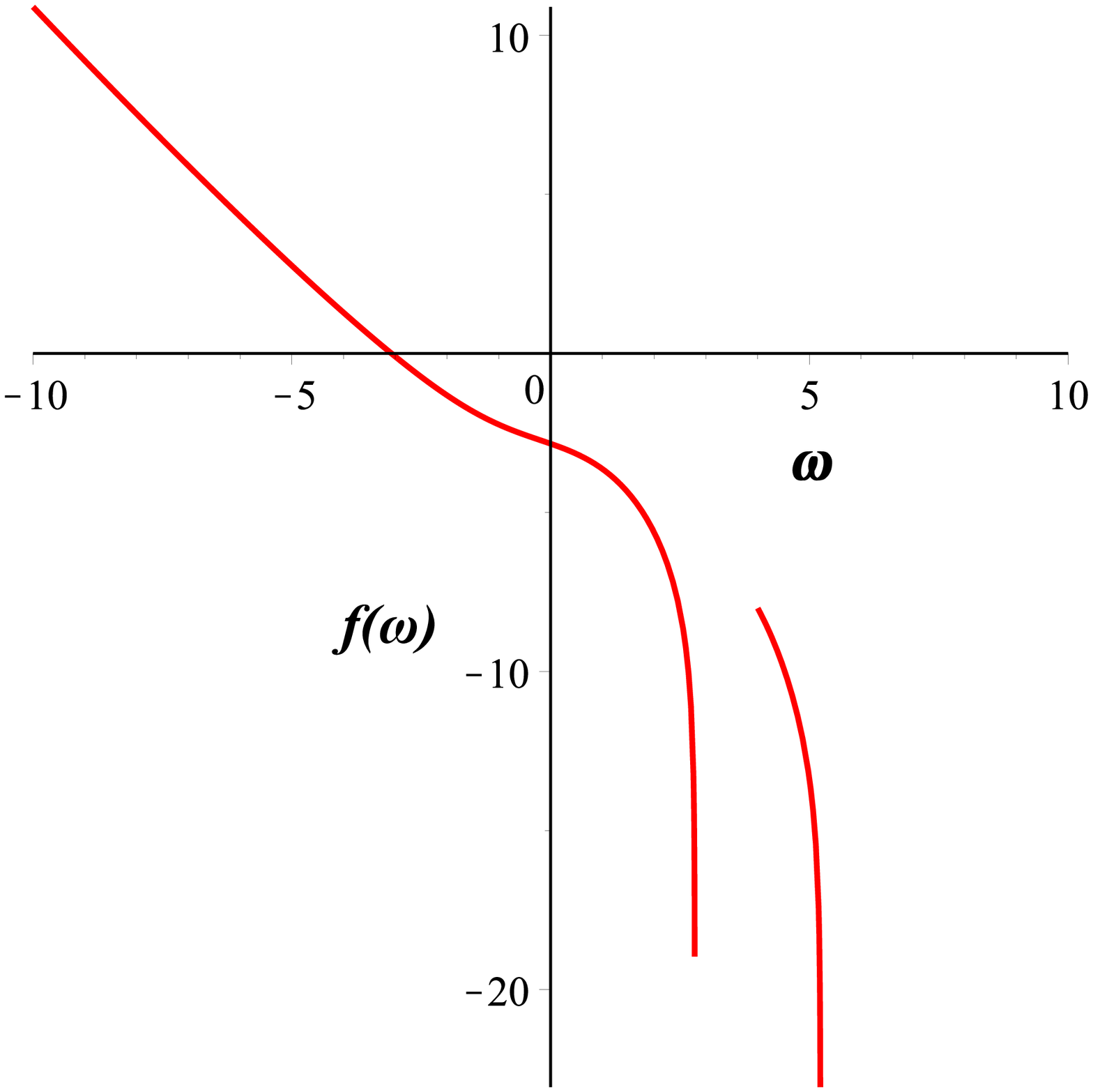}
		\caption{The shape function of the blue noise for the parameter set
			$a =  \lambda = \nu = 1$  and
			initial conditions $c_1 =1 $ and $ c_2 = 1$, respectively.}
		\label{blue}       
		
		\includegraphics[keepaspectratio, width=7cm]{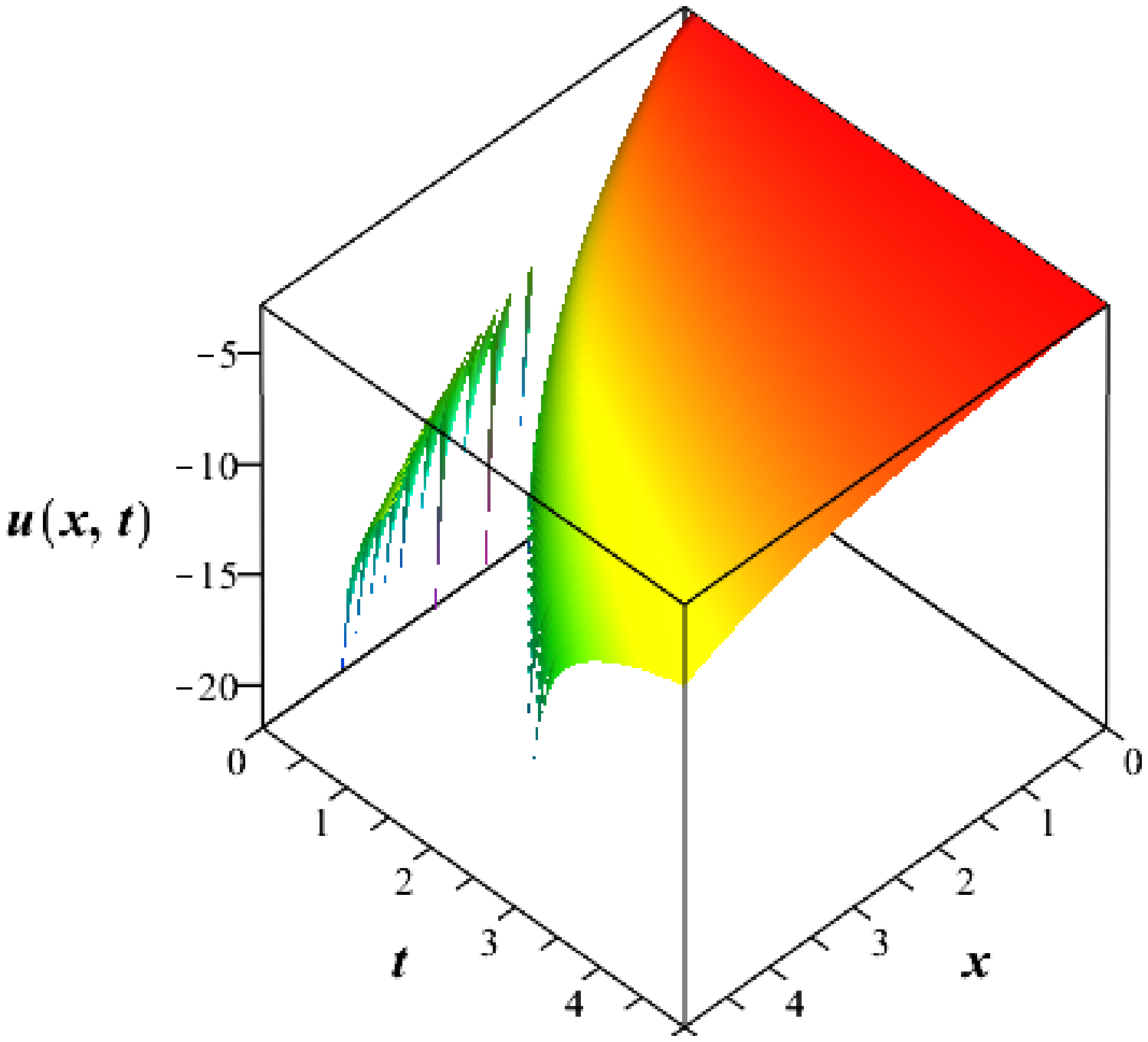}
		\caption{The complete solution $u(x,t)$ for the blue noise with the
			parameters given above.}
		\label{blue_3d}       
	\end{center}
\end{figure}

\subsection{Gaussian noise}

The first non-power law noise gives us the ODE of
\begin{equation}
\nu f''(\omega) + f'(\omega)\left [\frac{\omega}{2}+
\frac{\lambda}{2} f'(\omega)\right]+ ae^{\frac{-\omega^2}{n}} = 0.
\label{kpz_gauss}
\end{equation}

There is no general formula available for arbitrary parameters
$\lambda, \eta, a$  and $n$. Fortunately, if two parameters are
fixed e.g.  $\nu = 1/2$ and $n=1$,  than there is a closed
expression available for the solution
\begin{equation}
f(\omega) =  -\frac{1}{2\lambda}  \ln\left[1+\tan \left\{\sqrt{
	\lambda  a \pi}  \cdot erf( \sqrt{ \frac{\omega}{2}} )+c_1 \right\}^2
\right] + c_2, \label{gauss_sol}
\end{equation}
where $erf$ means the error function \cite{NIST}. Figure 12 presents
various solutions for different  values of $\lambda$. The larger the
value $\lambda$ an the smaller the  parameter $a$ the more the number of initial islands are,
which is a remarkable new feature. The solution itself is a continuous function on the whole $\omega$ axis.
The Van Hove singularities at finite $\omega$  are still present.
Note, that at larger  value of $\lambda$, the depth of the singularity valleys become shallower.

Figure 13. presents the final solution of the PDE.  The general
features are very similar to the formerly investigated  noise
$a/\omega^2$ but now three independent islands increases as time
goes on. The islands never grow together,  the valleys stay present
even at large time.

At this point we mention, that for the exponential distribution
$\eta = e^{-\omega/a}$ as noise term, there is no analytic solution
available at all.

\begin{figure}
	\begin{center}
		\includegraphics[keepaspectratio, width=7cm]{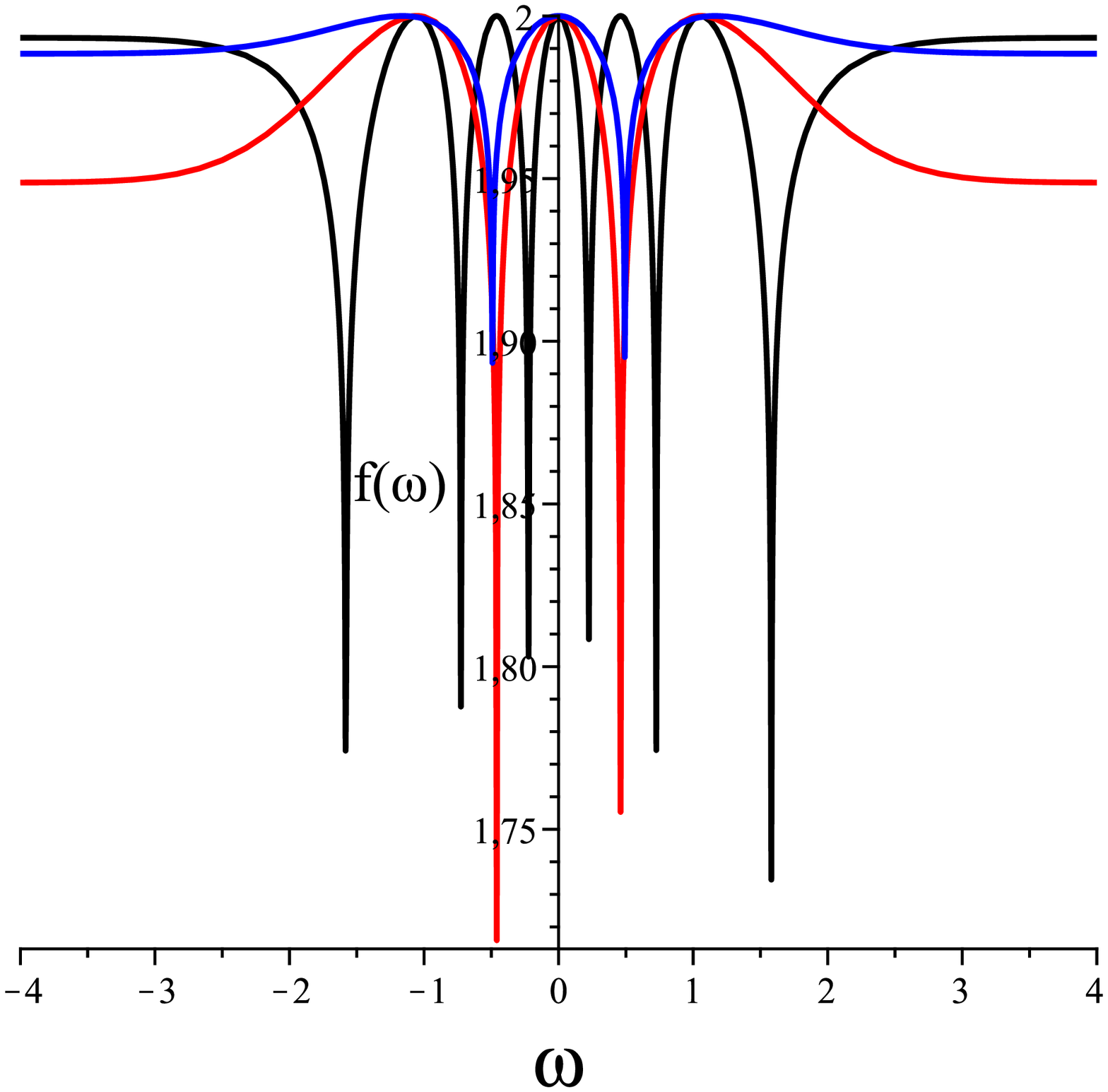}
		\caption{Various shape functions of Eq.(\ref{gauss_sol}) for the
			parameter set   $  \nu = 1/2,  n =1,   c_2 =1  $ and $c_1 =0$ for
			different $\lambda$ values. Black, red and blue lines are for $ a=
			1, \lambda = 25$,                  $ a = 0.25, \lambda = 25$, and $
			a = 0.1, \lambda = 55$, respectively. }
		\label{hetedgs}       
		
		\includegraphics[keepaspectratio, width=7cm]{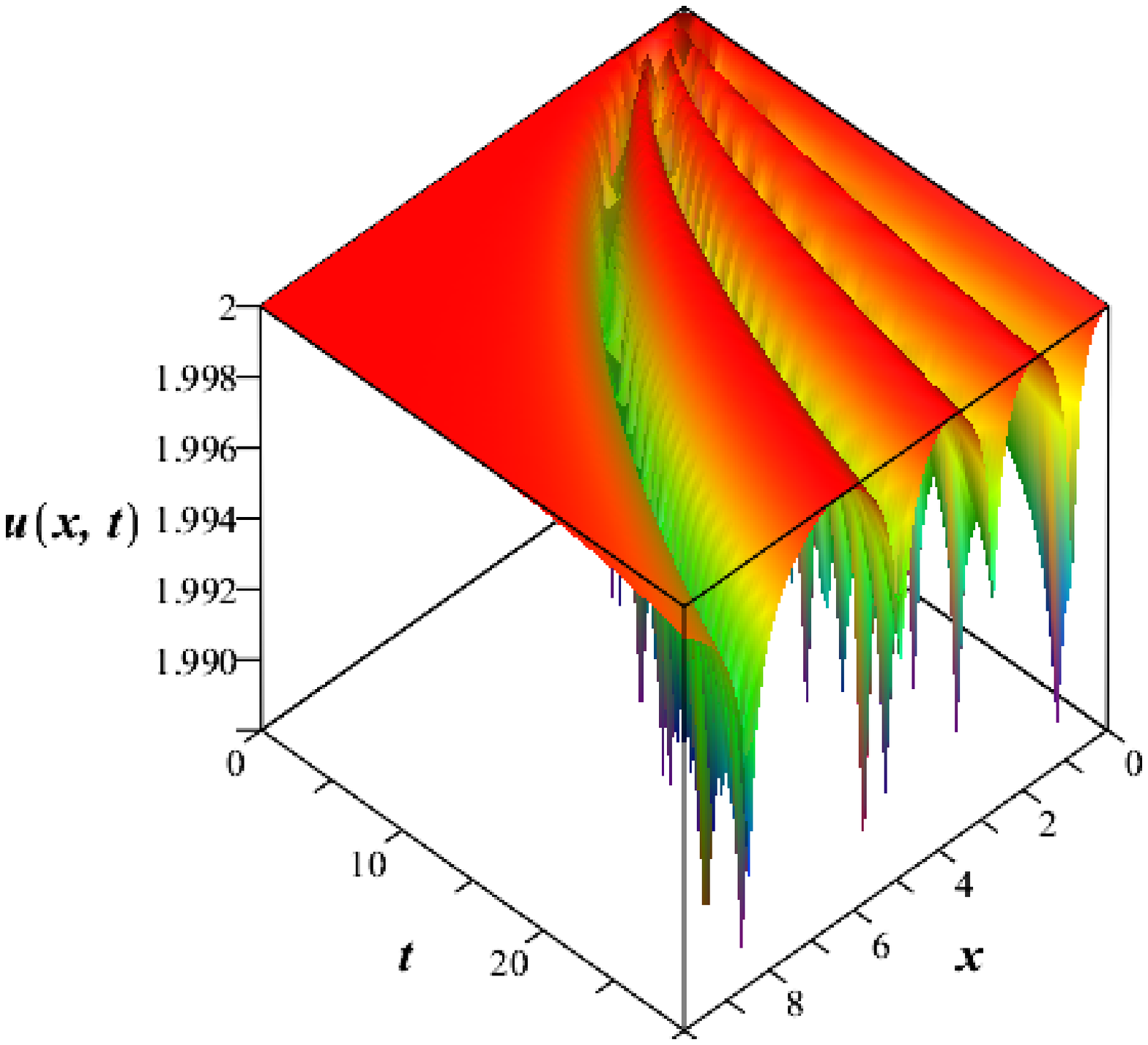}
		\caption{The complete solution for the Gaussian noise for $a = 0.1,
			\lambda = 55$ value. Other parameters are unchanged. }
		\label{hfdgetes}       
	\end{center}
\end{figure}

\subsection{Lorenzian noise}
As last system we have to investigate the ODE of
\begin{equation}
\nu f''(\omega) + f'(\omega)\left [\frac{\omega}{2}+
\frac{\lambda}{2} f'(\omega)\right]+ \frac{a}{1+\omega^2}= 0.
\label{kpz_lorenz}
\end{equation}

Unfortunately, the general solution cannot again be given in a
closed form. In the formal solution some integrals of the Heun
functions remain. For positive and given parameters $\nu, \lambda$,
the solution becomes well-defined. As an example for $ a=1/2,
\lambda = 2$ and $\nu =1/2$, the shape function reads
\begin{eqnarray}
f(\omega) &=&  -\frac{\omega^2} {4}+  \frac{\ln(2)}{2}  + \frac{1}{2}\ln \left\{
\left[ -c_2 \omega H_C\left(-\frac{1}{2}, \frac{1}{2}, 1, \frac{1}{8}, \frac{7}{8}, -\omega^2\right)     +   \right. \right. \nonumber \\
&& \left. \left.  c_1H_C \left(-\frac{1}{2}, -\frac{1}{2}, 1, \frac{1}{8}, \frac{7}{8}, -\omega^2\right) \right]/
\right. \nonumber \\
&&     \left[    -2 \omega^4 H_C\left(-\frac{1}{2},- \frac{1}{2}, 1, \frac{1}{8},
\frac{7}{8}, -\omega^2 \right) H_{CPrime}\left(-\frac{1}{2},  \frac{1}{2}, 1, \frac{1}{8}, \frac{7}{8}, -\omega^2 \right)  +  \right. \nonumber \\
&& \left.  2\omega^4  H_{CPrime}\left(-  \frac{1}{2},- \frac{1}{2}, 1, \frac{1}{8}, \frac{7}{8}, -\omega^2    \right)
H_C\left(- \frac{1}{2},\frac{1}{2}, 1, \frac{1}{8}, \frac{7}{8}, -\omega^2 \right) +  \right.   \nonumber \\
&&\omega^2  H_C\left(-\frac{1}{2},- \frac{1}{2}, 1, \frac{1}{8}, \frac{7}{8}, -\omega^2 \right)
H_C\left(- \frac{1}{2}, \frac{1}{2}, 1, \frac{1}{8}, \frac{7}{8}, -\omega^2\right)    + \nonumber \\
&& 2\omega^2 H_{CPrime} \left(-  \frac{1}{2},- \frac{1}{2}, 1, \frac{1}{8}, \frac{7}{8}, -\omega^2\right)
H_C \left( - \frac{1}{2}, \frac{1}{2}, 1, \frac{1}{8}, \frac{7}{8}, -\omega^2 \right)  -  \nonumber \\
&& 2\omega^2 H_{C} \left( -\frac{1}{2},- \frac{1}{2}, 1, \frac{1}{8}, \frac{7}{8}, -\omega^2 \right)
H_{CPrime} \left(- \frac{1}{2}, \frac{1}{2}, 1, \frac{1}{8}, \frac{7}{8}, -\omega^2  \right)   +  \nonumber  \\
&& \left.  \left.   H_C\left( -\frac{1}{2},- \frac{1}{2}, 1, \frac{1}{8}, \frac{7}{8}, -\omega^2 \right)
H_C\left(- \frac{1}{2}, \frac{1}{2}, 1, \frac{1}{8}, \frac{7}{8}, -\omega^2  \right),     \right]  \right\}
\label{lorentz}
\end{eqnarray}
where $H_C$ and $H_{CPrime}$  means the Heun functions and the
derivative of the Heun $C$ function, respectively \cite{NIST}.
\begin{figure}
	\begin{center}
		\includegraphics[keepaspectratio, width=7cm]{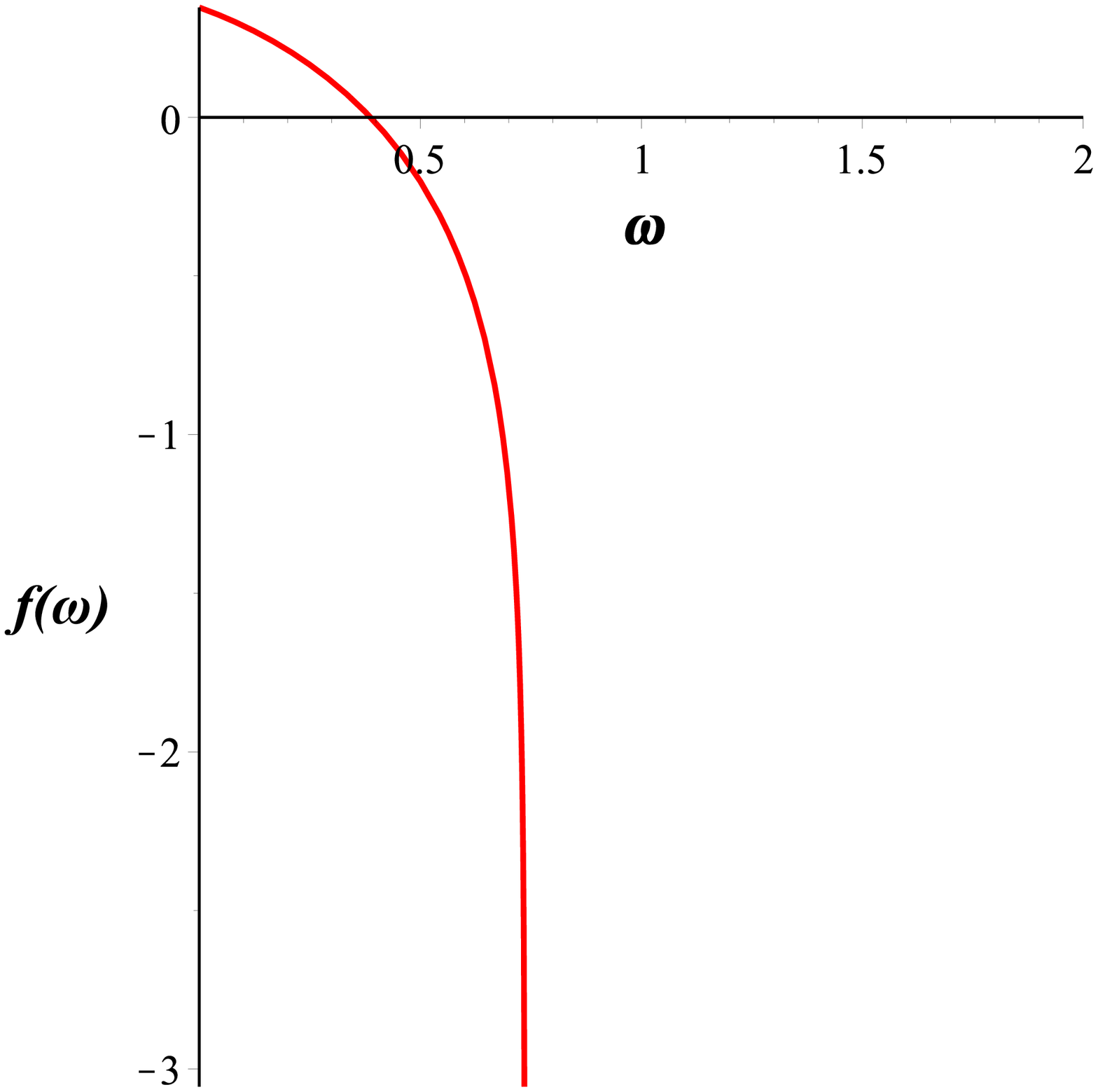}
		\caption{The shape function for the Lorenz noise for $a= 1/2, \lambda = 2$, $\nu =
			1/2$ and for
			$c_1 = c_2 = 1.$  }
		\label{Lorenz}       
		
		\includegraphics[keepaspectratio, width=7cm]{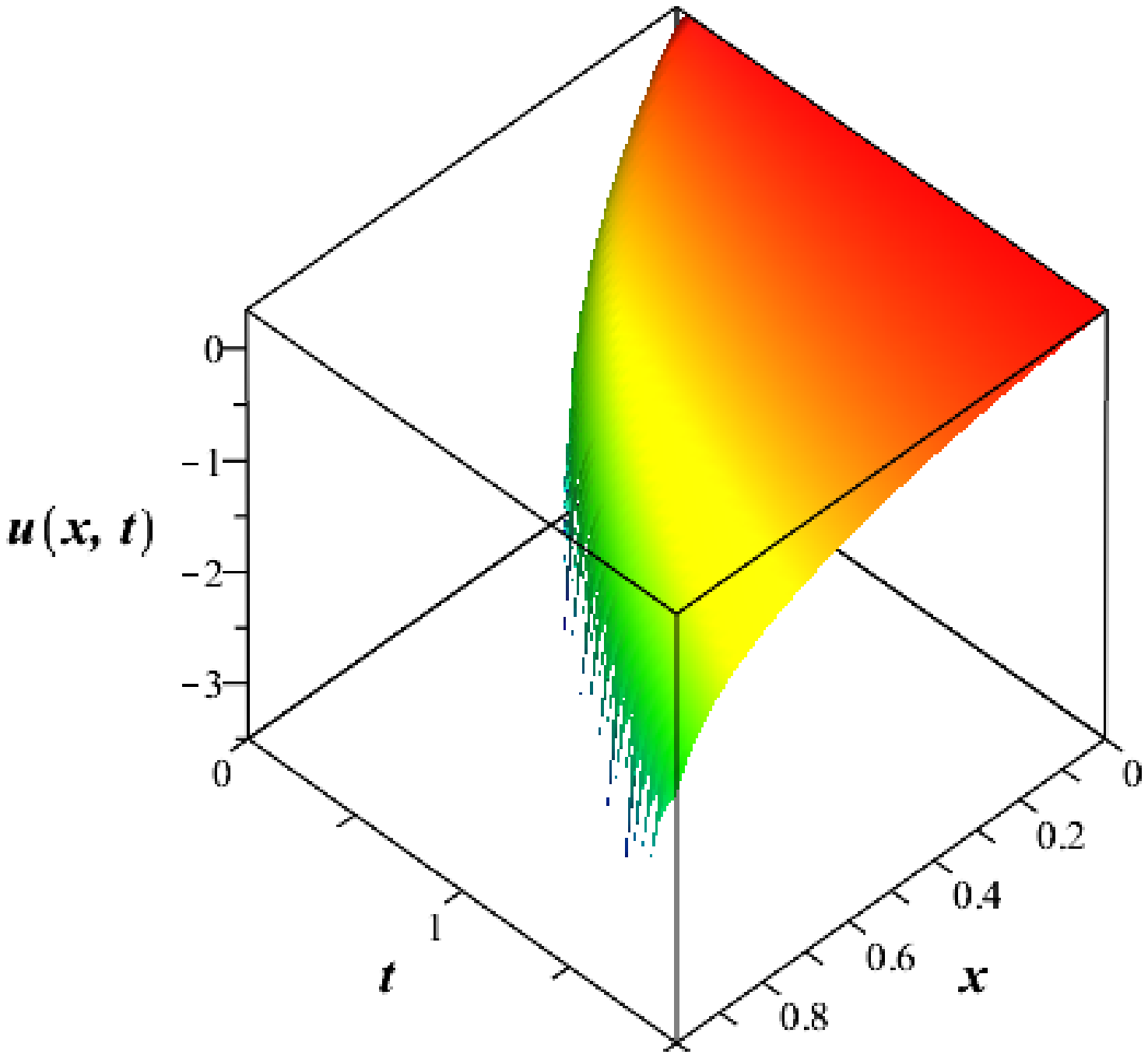}
		\caption{The complete solution  $u(x,t)$ for the Lorenz noise with the parameters given above.}
		\label{Lorenz_3D}       
	\end{center}
\end{figure}
Figure 14 shows the shape function for the Lorenzian noise term. The new
feature compared to the former Gaussian noise term is that the domain of the solution is just  a finite interval. Just a single island is born at the beginning of the surface growth process.
The solution blows up (or blows down) on a finite one is the compact support of the
solution. The last figure (Fig. 15) presents the final solution of the KPZ PDE. The
solution has a compact support as well. It means that the small
island which was positioned at the origin just grows for large
times, but cannot diffuse onto the whole surface.


\section*{Conclusions}
In summary we can say that with an appropriate change of variables
applying the self-similar Ansatz one may obtain analytic solution
for the KPZ equation for one spatial dimension with numerous noise
terms. We investigated four power-law-type  noise $\omega^n$ with
exponents of $-2,-1,0,1$, called the brown, pink, white and blue
noise, respectively. Each integer exponent describes completely
different dynamics. Additionally, we investigated the properties of
Gaussian and Lorenzian noises. Providing completely dissimilar
surfaces with growth dynamics. All solutions can be described with
non-trivial combinations of various special functions, like error,
Whittaker, Kummer or Heun. The parameter dependencies of the
solutions are investigated and discussed. Future works are planned
for the investigations of the two dimensional surfaces.

We also remark that applying transformations $k=e^{\frac{\lambda}{2
		\nu }u}$, $k=t^{\alpha}m(z)$ and $z =xt^{-\beta}$ to equation
(\ref{kpz}), one gets the linear ordinary differential equation
$m''+\frac{1}{2}zm'+(\frac{\lambda}{2 \nu }t\eta-\alpha )m=0$ for
any arbitrary value of $\alpha $ and $\beta =1/2$.

\section*{Acknowledgment}

This work was supported by Project no. 129257  implemented with the
support provided from the National Research, Development and
Innovation Fund of Hungary, financed under the $K\_  18$ funding
scheme.

\bibliographystyle{unsrt}  
\bibliography{references}  

\newcommand{\nosort}[1]{}
\begin{thebibliography}{10}

\bibitem{konyv}
A.~Pimpinelli and J.~Villain.
\newblock {\em Physics of Crystal Growth}.
\newblock Cambridge University Press, 1998.

\bibitem{kpz}
G.~Parisi M.~Kardar and Yi-C. Zhang.
\newblock Dynamic scaling of growing interfaces.
\newblock {\em Phys. Rev. Lett.}, 56:889, 1986.

\bibitem{barab}
A.-L. Barab{\'a}si and H.~E. Stanley.
\newblock {\em Fractal concepts in surface growth}.
\newblock Press Syndicate of the University of Cambridge, 1995.

\bibitem{hwa1}
T.~Hwa and E.~Frey.
\newblock Exact scaling function of interface growth dynamics.
\newblock {\em Physical Review A}, 44(12):R7873, 1991.

\bibitem{hwa2}
E.~Frey, U.~C. T{\"a}uber, and T.~Hwa.
\newblock Mode-coupling and renormalization group results for the noisy burgers
  equation.
\newblock {\em Physical Review E}, 53(5):4424, 1996.

\bibitem{lass}
M.~L{\"a}ssig.
\newblock On growth, disorder, and field theory.
\newblock {\em Journal of physics: Condensed matter}, 10(44):9905, 1998.

\bibitem{krug}
T.~Kriecherbauer and J.~Krug.
\newblock A pedestrian's view on interacting particle systems, kpz universality
  and random matrices.
\newblock {\em Journal of Physics A: Mathematical and Theoretical},
  43(40):403001, 2010.

\bibitem{einax}
M.~Einax, W.~Dieterich, and P.~Maass.
\newblock Colloquium: Cluster growth on surfaces: Densities, size
  distributions, and morphologies.
\newblock {\em Reviews of modern physics}, 85(3):921, 2013.

\bibitem{Matsushita}
M.~Matsushita, J.~Wakita, H.~Itoh, I.~Rafols, T.~Matsuyama, H.~Sakaguchi, and
  M.~Mimura.
\newblock Interface growth and pattern formation in bacterial colonies.
\newblock {\em Physica A: Statistical Mechanics and its Applications},
  249(1-4):517--524, 1998.

\bibitem{kuram1}
Y.~Kuramoto and T.~Tsuzuki.
\newblock Persistent propagation of concentration waves in dissipative media
  far from thermal equilibrium.
\newblock {\em Progress of theoretical physics}, 55(2):356--369, 1976.

\bibitem{kuram2}
G.~I. Sivashinsky.
\newblock Large cells in nonlinear marangoni convection.
\newblock {\em Physica D: Nonlinear Phenomena}, 4(2):227--235, 1982.

\bibitem{guedda}
M.~Guedda and R.~Kersner.
\newblock Self-similar solutions to the generalized deterministic kpz equation.
\newblock {\em Nonlinear Differential Equations and Applications NoDEA},
  10(1):1--13, 2003.

\bibitem{kersner}
R.~Kersner and M.~Vicsek.
\newblock Travelling waves and dynamic scaling in a singular interface
  equation: analytic results.
\newblock {\em Journal of Physics A: Mathematical and General},
  30(7):2457--2465, 1997.

\bibitem{odor}
J.~Kelling, G.~{\'O}dor, and S.~Gemming.
\newblock Suppressing correlations in massively parallel simulations of lattice
  models.
\newblock {\em Computer Physics Communications}, 220:205--211, 2017.

\bibitem{mart}
T.~Martynec and S.~H.~L. Klapp.
\newblock Impact of anisotropic interactions on nonequilibrium cluster growth
  at surfaces.
\newblock {\em Phys. Rev. E}, 98:042801, Oct 2018.

\bibitem{sergi}
D.~Sergi, A.~Camarano, J.~M. Molina, A.~Ortona, and J.~Narciso.
\newblock Surface growth for molten silicon infiltration into carbon
  millimeter-sized channels: Lattice--boltzmann simulations, experiments and
  models.
\newblock {\em International Journal of Modern Physics C}, 27(06):1650062,
  2016.

\bibitem{melo}
B.~A. Mello.
\newblock A random rule model of surface growth.
\newblock {\em Physica A: Statistical Mechanics and its Applications},
  419:762--767, 2015.

\bibitem{CaDoRo10}
P.~Calabrese, P.~L. Doussal, and A.~Rosso.
\newblock Free-energy distribution of the directed polymer at high temperature.
\newblock {\em EPL (Europhysics Letters)}, 90(2):20002, 2010.

\bibitem{SaSp10}
T.~Sasamoto and H.~Spohn.
\newblock One-dimensional kardar-parisi-zhang equation: an exact solution and
  its universality.
\newblock {\em Physical review letters}, 104(23):230602, 2010.

\bibitem{CaDo11}
P.~Calabrese and P.~L. Doussal.
\newblock Exact solution for the kardar-parisi-zhang equation with flat initial
  conditions.
\newblock {\em Physical review letters}, 106(25):250603, 2011.

\bibitem{DoTh17}
P.~L. Doussal and T.~Thiery.
\newblock Diffusion in time-dependent random media and the kardar-parisi-zhang
  equation.
\newblock {\em Physical Review E}, 96(1):010102, 2017.

\bibitem{MaGa05}
L.~M{\'a}ty{\'a}s and P.~Gaspard.
\newblock Entropy production in diffusion-reaction systems: The reactive random
  lorentz gas.
\newblock {\em Physical Review E}, 71(3):036147, 2005.

\bibitem{MaTeVo01}
L.~M{\'a}ty{\'a}s, T.~T{\'e}l, and J.~Vollmer.
\newblock Multibaker map for shear flow and viscous heating.

\bibitem{barna2016self}
I.~F. Barna, G.~Bogn\'ar, and K.~Hricz\'o.
\newblock Self-similar analytic solution of the two-dimensional navier-stokes
  equation with a non-newtonian type of viscosity.
\newblock {\em Mathematical Modelling and Analysis}, 21(1):83--94, 2016.

\bibitem{barn}
I.~F. Barna and R.~Kersner.
\newblock Heat conduction: a telegraph-type model with self-similar behavior of
  solutions.
\newblock {\em Journal of Physics A: Mathematical and Theoretical},
  43(37):375210, 2010.

\bibitem{sedov}
L.~I. Sedov.
\newblock {\em Similarity and dimensional methods in mechanics}.
\newblock CRC press, 1993.

\bibitem{barna2013analytic}
I.~F. Barna and L.~M\'aty\'as.
\newblock Analytic solutions for the one-dimensional compressible euler
  equation with heat conduction and with different kind of equations of state.
\newblock {\em Miskolc Mathematical Notes}, 14(3):785--799, 2013.

\bibitem{barna2}
I.~F. Barna.
\newblock Self-similar solutions of three-dimensional navier—stokes equation.
\newblock {\em Communications in Theoretical Physics}, 56(4):745, 2011.

\bibitem{barna3}
I.~F. Barna.
\newblock Self-similar analysis of various navier-stokes equations in two or
  three dimensions.
\newblock In D.~Campos, editor, {\em Handbook on Navier-Stokes Equations},
  Theory and Applied Analysis, New York, 2017. Nova Publishers.

\bibitem{NIST}
W.~J.~F. Olver, D.~W. Lozier, R.~F. Boisvert, and C.~W. Clark.
\newblock {\em NIST handbook of mathematical functions hardback and CD-ROM}.
\newblock Cambridge University Press, 2010.

\bibitem{vanhove}
L.~Van Hove.
\newblock The occurrence of singularities in the elastic frequency distribution
  of a crystal.
\newblock {\em Physical Review}, 89(6):1189, 1953.

\bibitem{vicsek}
Z.~Csah{\'o}k, K.~Honda, E.~Somfai, M.~Vicsek, and T.~Vicsek.
\newblock Dynamics of surface roughening in disordered media.
\newblock {\em Physica A: Statistical Mechanics and its Applications},
  200(1-4):136--154, 1993.

\end{thebibliography}






\end{document}